\documentclass[reprint,aps,showpacs,amssymb,amsfonts,superscriptaddress,twocolumn,prl,nofootinbib]{revtex4-2}

\usepackage{bbm,mathrsfs}
\usepackage{graphicx}
\usepackage{amsfonts,amsmath}
\usepackage{amsthm}
\usepackage{times}
\usepackage{longtable}
\usepackage{color}
\usepackage[colorlinks=true,citecolor=blue,urlcolor=blue]{hyperref}

\def\textbf#1{{\bf #1}}
\def\be{\begin{equation}}
	\def\ee{\end{equation}}
\def\ben{\begin{eqnarray}}
	\def\een{\end{eqnarray}}
\def\eea{\end{array}}
\def\bea{\begin{array}}
\def\bma{\begin{bmatrix}}
	\def\ema{\end{bmatrix}}

\newcommand{\bei}{\begin{itemize}}
	\newcommand{\eei}{\end{itemize}}
\newcommand{\ket}[1]{|#1\rangle}
\newcommand{\bra}[1]{\langle#1|}

\definecolor{jordi}{rgb}{0.1,0.1,0.5}

\usepackage{changes}

\definecolor{myurlcolor}{rgb}{0,0,0.7}
\definecolor{myrefcolor}{rgb}{0.8,0,0}
\usepackage{hyperref}
\hypersetup{colorlinks, linkcolor=myrefcolor,
	citecolor=myurlcolor, urlcolor=myurlcolor}

\theoremstyle{definition}

\usepackage[capitalise]{cleveref}
\definecolor{OliveGreen}{rgb}{0,0.6,0}

\begin{document}
	
	\title{Tropical contraction of tensor networks as a Bell inequality optimization toolset}
	
	\author{Mengyao Hu}
	\email{mengyao@lorentz.leidenuniv.nl}
	\affiliation{Instituut-Lorentz, Universiteit Leiden, P.O. Box 9506, 2300 RA Leiden, The Netherlands}
	\author{Jordi Tura}
	\email{tura@lorentz.leidenuniv.nl}
	\affiliation{Instituut-Lorentz, Universiteit Leiden, P.O. Box 9506, 2300 RA Leiden, The Netherlands}

	\begin{abstract}
		We show that finding the classical bound of broad families of Bell inequalities can be naturally framed as the contraction of an associated tensor network, but in tropical algebra, where the sum is replaced by the minimum and the product is replaced by the arithmetic addition. 
        We illustrate our method with paradigmatic examples both in the multipartite scenario and the bipartite scenario with multiple outcomes. We showcase how the method extends into the thermodynamic limit for some translationally invariant systems and establish a connection between the notions of tropical eigenvalue and the classical bound per particle as a fixed point of a tropical renormalization procedure.
	\end{abstract}
	
	\keywords{} \pacs{}

	\maketitle
	
	\paragraph{Introduction} ---
	\label{sec:intro}
	Nonlocality~\cite{BrunnerRMP2014} is one of the most striking features of quantum physics. 
    Since its inception, intense debates among its founders took place~\cite{EPR35}. 
    Eventually settled by Bell with the formal introduction of a local hidden variable model (LHVM)~\cite{Bell1964}, and later demonstrated by experiments~\cite{AspectPRL82, HensenNature2015, GiustinaPRL2015, ShalmPRL2015, AbellanNature2018}, nonlocality is established as the impossibility for a set of correlations to be explainable within the paradigm of local realism. 
    Nonlocality is detected via the violation of a so-called Bell inequality, which geometrically corresponds to a half-space containing all LHVM correlations. 
    Beyond its importance in quantum foundations, nonlocality is the key resource for device-independent (DI) quantum information processing (QIP), enabling tasks such as DI quantum key distribution~\cite{AcinDIQKD, PironioPRX2013}, DI randomness expansion~\cite{PhDColbeck} and amplification~\cite{ColbeckRenner2012, GallegoNatComms2013}, or DI self-testing~\cite{SupicQuantum2020}.
	
	The advent of DIQIP, as well as continued theoretical and experimental advances, has motivated the search for Bell inequalities way beyond the pioneering Clauser-Horne-Shimony-Holt (CHSH) inequality~\cite{CHSH}. 
    The CHSH inequality belongs to the only non-trivial class of tight Bell inequalities for the simplest bipartite scenario~\cite{Faacets}, where each party can perform one out of two dichotomic measurements.
	In the multipartite case, however, finding all Bell inequalities simply becomes an intractable task~\cite{Pitowski1989, PitowskiSvozilPRA2001, ChazelleDCG1993, GuehnePRL2005, TothPRA2006}. Nevertheless, any linear combination of observable correlators can be converted into a Bell inequality by adding an appropriate constant shift (the so-called classical bound) by solving a simpler --yet still intractable in general-- optimization task.

	Interestingly, finding the classical bound for a given Bell inequality is in one-to-one correspondence with finding the minimal energy configuration of an associated classical spin system~\cite{TuraPRX2017, Emonts2024Effects}, which is a well-established NP-hard problem~\cite{BarahonaJPA1982}. 
    For classical spin models on a one-dimensional geometry with finite range interactions, however, efficient algorithms exist for finding its ground state~\cite{SchuchPRA2010}, based on a dynamic programming (DP) approach.

	In this work, we show that such combinatorial optimization task for Bell inequalities consisting of $r$-body correlators, with $r<n$, can be formulated as contracting an associated tensor network, but in the so-called tropical algebra~\cite{MaclaganBook2015}, a natural mathematical framework for combinatorial optimization.
    This allows us to give a systematic method for optimizing broad classes of Bell inequalities, even reaching the thermodynamic limit by extending this notion to infinite, translationally invariant systems~\cite{WangPRL2017, yang2022contextuality}.

	Our work rests on two fundamental ideas: on the one hand, the tensor network (TN) formalism. TN methods have proven to be a tremendously successful approach for tackling many-body quantum systems~\cite{CiracRMP2021}, and for detecting nonlocality, e.g., in the context of connector TN~\cite{NavascuesPRX2018}.
    On the other hand, tropical algebra, which can be thought of as piecewise linear algebra, is the adequate framework for describing DP and many combinatorial optimization tasks~\cite{MaclaganBook2015}. 
    For instance, finding minimal energy configurations for classical spin models~\cite{LiuPRL2021} or even benchmarking quantum simulators for solving maximum independent set problems~\cite{EbadiScience2022, liu_computing_2023}.
	
	\paragraph{Preliminaries} ---
	Let us begin by defining the arithmetic operations of tropical addition $x\oplus y:= \min(x, y)$ and tropical multiplication $x \odot y:= x + y$. 
    These operations on the real numbers plus infinity give rise to the tropical semiring, also referred to as the min-sum algebra~\cite{MaclaganBook2015}. To motivate such a definition, let us consider the following function of $n$ variables:
	
	\begin{equation}
		H(\mathbf{x}) = \sum_i f_i(x_i,x_{i+1}), \qquad x_i \in S,
		\label{eq:H}
	\end{equation}
	where $S$ is a finite set of $|S|$ elements and $f_i:S\times S \longrightarrow {\mathbbm R}\cup \{\infty\}$. 
    For instance, if $S=\{-1,1\}$ and $f_i(x_i,x_{i+1}) = h_i x_i x_{i+1}$, then $H$ can be interpreted as a classical Hamiltonian in a one-dimensional geometry. 
    A central problem in both physics and optimization is to find the minimal value $\beta$ of $H$; \textit{i.e.,} $\beta:=\min_{\mathbf{x}} H(\mathbf{x})$.
	
	Observe that for the case of \cref{eq:H}, this multivariate optimization can be efficiently solved by successive elimination of variables~\cite{TuraPRX2017, SchuchPRA2010}: since $i$-th variable only affects $f_{i-1}$ and $f_i$, we can define a new function $g_{i-1}$ that is independent of $x_i$:
	\begin{equation}
		g_{i-1}(x_{i-1},x_{i+1}) := \min_{x_i \in S} f_{i-1}(x_{i-1},x_i)+f_i(x_i,x_{i+1}).
		\label{eq:def:g}
	\end{equation}
	In tropical algebra terms, \cref{eq:def:g} is equivalent to
	\begin{equation}
		g_{i-1}(x_{i-1},x_{i+1}) = \bigoplus_{x_i \in S} f_{i-1}(x_{i-1},x_i) \odot f_i(x_{i},x_{i+1});
	\end{equation}
	\textit{i.e.,} \cref{eq:def:g} is the tropical inner product of the vectors $f_{i-1}(x_{i-1},\cdot)$ and $f_{i}(\cdot,x_{i+1})$. 
    This motivates the following definition: to each $f_i$ we assign a $|S|\times |S|$ matrix
	\begin{equation}
		F_i := (f_i(x_i,x_{i+1}))_{x_i,x_{i+1}\in S}.
	\end{equation}
	Then, the matrix $G_{i-1}$ associated to $g_{i-1}$ is simply given as the tropical matrix multiplication of $F_{i-1}$ and $F_i$, which we denote
	\begin{equation}
		G_{i-1} = F_{i-1}\odot F_i.
		\label{eq:G}
	\end{equation}
	Clearly, this procedure can be repeated until only the first and last variables remain, converging to a function $z(x_0,x_{n-1})$, with an associated matrix $Z=\bigodot_i F_i$. 
    If the function $H$ in \cref{eq:H} is defined with open boundary conditions (OBC), then the last optimization step is carried for $x_0$ and $x_{n-1}$ independently as
	\begin{equation}
		\beta^{\mathrm{OBC}} = \min_{(x_0,x_{n-1})\in S \times S} z(x_0,x_{n-1}) \equiv \bigoplus_{ x_0,x_{n-1}\in S} z(x_0,x_{n-1}).
	\end{equation}
	On the other hand, if periodic boundary conditions (PBC) are considered, then the extremes must coincide, so that the minimum needs to be taken over the pairs $(x_0,x_{n-1})$ satisfying $x_0=x_{n-1}$:
	\begin{equation}
		\beta^{\mathrm{PBC}} = \min_{x\in S} z(x,x) \equiv \bigoplus_{ x\in S} z(x,x).
	\end{equation}
	
	One quickly observes that both cases can be expressed in terms of the tropical trace, by defining the tropical trace of a matrix to be the tropical addition of its diagonal elements. 
    This relation is obvious for the PBC case, whereas the OBC requires an additional step, so that we obtain a tropical sum over all the elements of $Z$. 
    In conventional arithmetic, this is achieved by taking the trace of $Z\cdot U$ where $U$ is a (rank one) matrix of ones, which are the neutral elements of the conventional multiplication. Hence, its tropical counterpart must use the neutral element of the tropical multiplication; \textit{i.e.,} a matrix of zeroes $\mathbf{0}$.
	Hence, we can express
	\begin{eqnarray}
		\left\{
		\begin{array}{lll}
			\beta^{\mathrm{OBC}}&=&\mathrm{tropTr}(Z\odot \mathbf{0}) = \bra{0}\odot Z \odot \ket{0}, \\
			\beta^{\mathrm{PBC}}&=&\mathrm{tropTr}(Z).
		\end{array}
		\right.
	\end{eqnarray}
	
\begin{figure}[h!]
		\centering
		\includegraphics{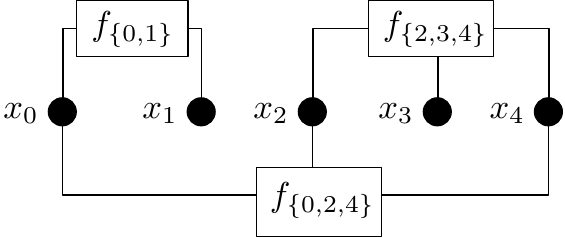}
		\caption{A graphical representation of the tropical contraction to minimize $H(x_0, x_1, x_2, x_3, x_4) = f_{\{0,1\}}(x_0, x_1) + f_{\{2,3,4\}}(x_2,x_3,x_4) + f_{\{0,2,4\}}(x_0,x_2,x_4)$ as $z(x_1,x_3)$.}
		\label{fig:TN}
	\end{figure}

	\paragraph{Tropical contraction of tensor networks} ---
	The example above motivates the introduction of a more general framework of tropical tensor networks and their contraction. 
    A more general function to consider is a sum of local terms; \textit{i.e.}, of the form
	\begin{equation}
		H(\mathbf{x}) = \sum_{I\in {\cal I}}f_I(\mathbf{x}_I),
		\label{eq:Hlocal}
	\end{equation}
	where $f_I(\mathbf{x}_I)$ depends only on the variables $x_i$ whose indices $i$ are in $I\subseteq [n] := \{0,\ldots, n-1\}$, with $|I| \in O(1)$ and $|{\mathcal I}|\in O(\mathrm{poly}(n))$.
	Each term $f_I(\mathbf{x}_I)$ is a function from $S^{|I|}$ to $\mathbbm{R}\cup \{\infty\}$ so it is specified by $|S|^{|I|}$ entries that can be arranged in a rank-$|I|$ tensor. 
    Diagrammatically, $F_{I}$ can be represented as a tensor with $|I|$ open indices, with the optimization of $H$ from \cref{eq:Hlocal} yielding the network contraction (e.g., \cref{fig:TN}).
	
	The optimization of \cref{eq:Hlocal} is therefore done via a contraction of a network like in \cref{fig:TN} in tropical algebra. 
    Let us formally define the elements involved. It is convenient to start with the tropical analogue of the Kronecker delta function
	\begin{equation}
		\delta_{\mathrm{trop}}(x,y) = \left\{
		\begin{array}{lll}
			0&\mbox{if}&x=y  \\
			\infty& \mbox{else}
		\end{array}.
		\right.
	\end{equation}
	This definition naturally extends to the multivariate Kronecker delta since, in tropical algebra,
	\begin{equation}
		\delta_{\mathrm{trop}}(\mathbf{x}) = \delta_{\mathrm{trop}}(x_0,x_1) \odot \cdots \odot \delta_{\mathrm{trop}}(x_{n-2},x_{n-1}).
		\label{eq:tropidtensor}
	\end{equation}
	Note that $\delta_{\mathrm{trop}}(\mathbf{x})$ is $0$ if, and only if, all the coordinates in $\mathbf{x}$ are equal; it is infinity otherwise. Hence, the $|S|$-dimensional tropical identity matrix is simply
	\begin{equation}
		\mathbbm{1}_{\mathrm{trop}} = \left(
		\begin{array}{cccc}
			0&\infty&\cdots&\infty\\
			\infty&\ddots&\ddots&\vdots\\
			\vdots& \ddots& \ddots& \infty\\
			\infty&\cdots& \infty &0
		\end{array}
		\right) = (\delta_{\mathrm{trop}}(x,y))_{x,y \in S}.
	\end{equation}
	
\begin{figure}[h!]
		\centering
		\includegraphics{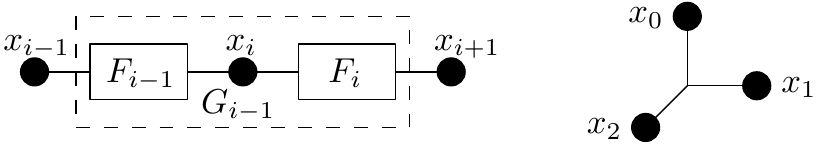}
		\caption{A graphical representation of $G_{i-1}=F_{i-1}\odot F_i$ (left) and $\delta_{\mathrm{trop}}(x_0,x_1,x_2)$ (right).}
		\label{fig:TN2}
	\end{figure}

	In \cref{fig:TN2} we display the contraction step introduced in \cref{eq:G} and the Kronecker delta \cref{eq:tropidtensor}.
	
	\paragraph{Contraction procedure} ---
	In order to contract a whole network, denoted by the set of indices ${\cal I}$ and the associated functions $f_I(\mathbf{x}_I)$ as $\{{\cal I}, \{f_I(\mathbf{x}_I)\}_{I \in {\cal I}}\}$, one needs to pick an elimination order on the $x_i$ that do not correspond to open indices. 
    If there are no open indices, the network contraction, denoted $\mathrm{tropC}({\cal I};\{f_I\}_{I\in {\cal I}})$ ends up in a scalar $\beta$, which is a zero-dimensional tensor. In $1\mathrm{D}$ networks one can iteratively eliminate variables $x_1,x_2, \ldots$ but this approach does not need to be the best way. 
    By eliminating e.g. the odd-site variables one can parallelize the network contraction and achieve exponential speedup in translationally invariant cases~\cite{TuraPRX2017}. 
    Picking an optimal contraction order for a tensor network is a very hard problem in general~\cite{EvenblyPRB2014, SchindlerMLSciTech2020} and one typically resorts to heuristics.

	Let us define the set of variables we want to eliminate
	\begin{equation}
		\label{eq:eli}
		\mathrm{Eli}({\mathcal I}) := \{x_i: \exists\ I, J \in {\mathcal I}, I\neq J\ \mbox{s.t. } x_i \in I \cap J\},
	\end{equation}
	which correspond to the closed indices of the network.
	Then, an elimination order is a permutation of $\mathrm{Eli}({\mathcal I})$ denoted by $\sigma \in {\mathfrak S}_{|\mathrm{Eli}({\mathcal I})|}$. At step $k>0$ we pick the $x_{\sigma(k)}$ to contract over. To this end, it is convenient to label the set $\cal I$ explicitizing the contraction step, so we denote ${\cal I}^{(0)}\equiv {\cal I}$ and $f_I^{(0)}\equiv f_I \ \forall I \in {\mathcal I}$. 
    To eliminate a variable $x_{\sigma(k)}$, we first identify all terms in the sum affected by it.

    Our goal will be to describe how to update ${\cal I}^{(k)}$ and $\{f_{I}^{(k)}\}_{I \in {\cal I}^{(k)}}$ at every step $k$.
	
	Let $\mathcal{J}_k \subseteq {\mathcal I}^{(k-1)}$ be the collection of indices in the sum that are affected by $x_{\sigma(k)}$; \textit{i.e.}, the neighborhood of $x_{\sigma(k)}$ defined by the network topology:
	\begin{equation}
		\label{eq:j_k}
		\mathcal{J}_k:=\{I\in {\mathcal I}^{(k-1)}:\ x_{\sigma(k)} \in I\}.
	\end{equation}
	Within the subnetwork defined by $\mathcal{J}_k$, we can identify those variables that can also be eliminated at the $k$-th step as the tensor legs that are closed indices within $\mathcal{J}_k$ and not connected to the rest of the network:
	\begin{equation*}
	    C_k:=\{x_i:\ \exists I\neq J \in {\mathcal J}_k, \ \mbox{s.t.}\ x_i \in I \cap J\}\setminus \bigcup_{I\in {\mathcal I}^{(k-1)}\setminus {\mathcal J}^{(k)}} I.
	\end{equation*}
	Finally, the open indices of the remaining contraction are given by
	\begin{equation*}
	    O_k:=\bigcup_{J \in {\mathcal J}_k} J \setminus C_k;
	\end{equation*}
	\textit{i.e.}, all the variables that are nearest neighbors of $x_{\sigma(k)}$ except those that can be eliminated also at the $k$-th step.

\begin{figure}[h!]
		\centering
		\includegraphics{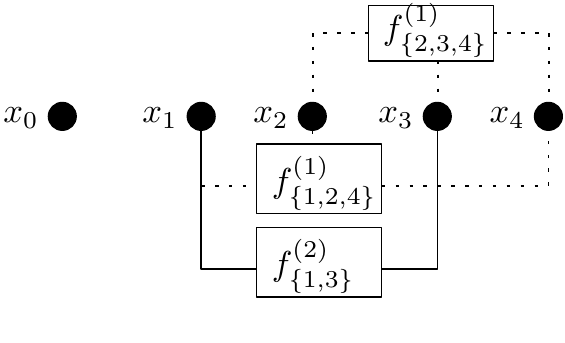}
		\caption{A graphical representation of the contraction of the example in \cref{fig:TN}, with $\mathrm{Eli}({\mathcal I})=\{x_0,x_2,x_4\}$ and elimination order $x_0, x_2, x_4$. Then, ${\mathcal J}_1=\{\{0,1\},\{0,2,4\}\}$, $C_1=\{0\}$, $O_1 = \{1,2,4\}$; ${\mathcal J}_2 =\{\{1,2,4\},\{2,3,4\}\}$, $C_2=\{2,4\}$ and $O_2=\{1,3\}$.}
			
		\label{fig:TN3}
	\end{figure}

	We can now update the network as $${\cal I}^{(k)}:=\{O_k\}\cup\{I\}_{I \in {\cal I}^{(k-1)}\setminus {\cal J}_{k}}$$
	and
	$$f_I^{(k)}(\mathbf{x}_{I}):= \left\{
	\begin{array}{lcl}
	f_I^{(k-1)}(\mathbf{x}_I)&\mbox{if}& I \in {\cal I}^{(k-1)}\setminus {\cal J}_{k},\\
	\displaystyle \bigoplus_{x_p \in C_k } \bigoplus_{x_{\sigma(k)}}\bigodot_{I' \in {\cal J}_k }f_{I'}(\mathbf{x}_{I'}) &\mbox{if}&I=O_k.
	\end{array}
	\right.$$
	This procedure and definitions are exemplified in Fig. \ref{fig:TN3}.

	The ordering $\sigma$ chosen in $\mathrm{Eli}({\mathcal I})$ can greatly affect the overall tropical contraction complexity. 
    Basically, one wants to avoid leaving too many open indices in intermediate steps of the contraction, as even storing a tensor with $m$ open indices requires $O(|S|^m)$ space.
    In the recent years, ever more sophisticated algorithms for tensor network contraction have been developed in the context of spoofing recent quantum inimitability experiments~\cite{AruteNature2019}, such as optimized heuristics for contraction order and index slicing~\cite{HuangNatComputSci2021, VillalongaNPJQI2019, RanBook2020,PanPRL2020, GrayQuantum2021}, which naturally map to the tropical algebra~\cite{EbadiScience2022, liu_computing_2023}.
	
	\paragraph{Tropical renormalization} ---
	In this section we focus on the translationally-invariant (TI), 1D case; \textit{i.e.,} \cref{eq:H} with all $f_i$ equal. 
    This case includes, but it is not limited to, all TI Bell inequalities on a $1$D geometry with correlators up to $O(1)$ neighbors~\cite{TuraPRX2017, TIpaper}, and serves as an entry point to Bell inequalities in hyperbolic or fractal geometries~\cite{Hu2024Characterizing}. 
    The study of the limiting case of these classes of inequalities as $n$ grows to infinity, reaching an infinite TI system in 1D~\cite{WangPRL2017} can be seen as an instance of tropical tensor renormalization.
	
	To this end, let us recall the graph interpretation of the tropical matrix product~\cite{MaclaganBook2015}: Let $F$ be the adjacency matrix of a directed graph $\mathfrak{g}(F)$, where $F_{ij}$ is the cost of going from node $i$ to node $j$. 
    If one considers $(F^{\odot 2})_{ij}$, that equals $\bigoplus_k F_{ik} \odot F_{kj} = \min_{k} F_{ik} + F_{kj}$; \textit{i.e.,} the minimal cost of going from node $i$ to node $j$ in exactly two steps, through an intermediate node indexed by $k$. 
    In general, $F^{\odot m}$ indexes the minimal costs of connecting two nodes in exactly $m$ steps and $\bigoplus_{k=1}^m F^{\odot k}$, in at most $m$ steps.
	
	Given the coefficients of a Bell inequality without full-body correlators, finding its classical bound can be viewed as a non-trivial instance of \cref{eq:Hlocal}, where $f_{I}(\mathbf{x}_I)$ corresponds to the sum of all the correlators that involve the parties in $I \subsetneq [n]$, and $S$ corresponds to the set of local deterministic strategies from which any party can choose. 
    Geometrically, this corresponds to a vertex of the local polytope (an element of $S^n$), and the optimal value of the classical bound can always be achieved at some vertex in $S^n$ as a consequence of the fundamental theorem of linear programming.
	
	In $1\mathrm{D}$ Bell inequalities with nearest-neighbor correlators, the optimization yielding the classical bound can be viewed as an instance of \cref{eq:H} and therefore carried out efficiently. 
    If the inequality has correlators that extend to the $r$-th nearest neighbor, then one first eliminates the parties whose index is a multiple of $r+1$ and effectively obtains a $1\mathrm{D}$ inequality with nearest neighbors, but now the set of parties in $\{(r+1)i+1, \ldots, (r+1)i+r\}$ is treated as a larger effective party, whose deterministic strategy is an element from $S^r$~\cite{TuraPRX2017}.

	Since $\mathfrak{g}(F)$ is strongly connected, as $F_{ij} < \infty$, there exists a unique tropical eigenvalue $\lambda(F)$~\cite{MaclaganBook2015}. A tropical eigenvalue-eigenvector pair $(\lambda, \mathbf{v})$ of $F$ satisfies $F\odot \mathbf{v} = \lambda \odot \mathbf{v}$ and $\lambda(F)$ can be understood as the minimum normalized cycle length in $\mathfrak{g}(F)$ if there are no cycles of negative weight~\cite{MaclaganBook2015}. 
    If we think in the thermodynamic limit, the system size will eventually be $n\geq |S|^r$. 
    Hence, at least two parties must choose the same local deterministic strategy, thereby forming a closed loop. 
    Therefore, globally optimal strategies must consist of a concatenation of loops with optimal normalized length. 
    To find these, one does not need to explore all cycles, but efficient algorithms exist, such as Bellman-Ford or Floyd-Warshall~\cite{MaclaganBook2015}. 
    Loosely speaking, the basic idea is to retain the $\arg\min$ information when performing the tensor tropical contraction and backtrack.
	
	Note that we can interpret $F_{ij}$ as the contribution to the classical bound where Alice chooses the $i$-th strategy and Bob chooses the $j$-th. 
    In addition, $(F^{\odot k})_{ij}$ is the sum of minimal costs when the left side chooses strategy $i$ and the right side chooses strategy $j$. Therefore, we see that the tropical eigenvalue of $F$ is strongly related to the classical bound.

    Clearly, a sufficient condition to reach a fixed point in the renormalization flow is that the sequence $\{F, F^{\odot 2}, \ldots, F^{\odot k}, \ldots\}$ stabilizes;
    \textit{i.e.,} there exist two integers $k_0$ and $\sigma$ for which
    $F^{\odot k+\sigma} = \lambda(F)^{\odot \sigma} \odot F^{\odot k},\quad \forall k \geq k_0$. Here $\sigma=\sigma(F)$ is known as the cyclicity of the matrix $F$~\cite{Nowak2014Tropical, Hu2024Characterizing}.
    In this case, the classical bound per particle converges to $\lambda$ at $O(1/n)$ speed when $k$ satisfies certain conditions.

	Analogous to the fact that in classical arithmetic a matrix $A^k$ tends to a projector onto its dominant eigenspace, whose eigenvalue can be extracted via its spectral radius $\rho(A) = \lim_{k\rightarrow \infty} ||A^k||^{1/k}$, we observe that the contribution per particle to the classical bound tends to a tropical spectral radius, which we can define as $\rho_{\mathrm{trop}}(F) := \lim_{k\rightarrow \infty} \mathrm{tropTr}(F^{\odot k})^{\odot 1/k}$. The tropical $k$-th root corresponds to divide by $k$ and the tropical trace of $F^{\odot k}$ looks for the minimum cost of a cycle of length exactly $k$. If $\{F, F^{\odot 2}, \ldots\}$ stabilizes, then $\beta_C/n \longrightarrow \lambda = \rho_{\mathrm{trop}}(F)$. 
    We explain this in detail in an extended version of this work in a companion article ~\cite{Hu2024Characterizing}, which includes studying the stabilization properties of the TI Bell inequalities,
    as well as using the tropical eigenvector to list all the optimal strategies such that the classical bound is attained.

    Let us illustrate the renormalization procedure with the inequality $I_G$ from~\cite{WangPRL2017}, defined as
	\begin{equation*}
		I_G:=\sum_i \sum_{k=0,1} c_k E_k^{(i)} + \sum_{r=1,2}\sum_{k,l=0,1} c_{k,l}E^{(i,i+r)}_{k,l} - \beta_C \geq 0,
	\end{equation*}
	where $E^{(i)}_k = \sum_{a=0,1}(-1)^aP_i(a|k)$ and $E^{(i,j)}_{k,l} = \sum_{a,b=0,1}(-1)^{a+b}P_{i,j}(ab|kl)$.
	Here, the inequality is for two inputs and two outputs, and next-nearest-neighbors. Hence, $|S|^2 = 16$ is the number of strategies from the effective nearest-neighbor inequality, so there is a tensor $F$ of dimensions $16\times 16$ defining $I_G$. 
    In this case, the sequence of $F$ stabilizes already at the second tropical power: $F\neq F^{\odot 2}$ but $F^{\odot 2} = F^{\odot 3}$ and, indeed, $\rho_{\mathrm{trop}}(F) = 6$, the same bound per particle given in~\cite{WangPRL2017} for a countably infinite system size.

	\paragraph{Multiple outcome Bell inequalities} ---
	Two notable classes of Bell inequalities for the bipartite scenario with $m$ observables, each yielding $d$ different outcomes, are the CGLMP~\cite{CollinsPRL2002} and SATWAP~\cite{SATWAP} inequalities. They both admit the form
	\begin{equation}
		I = \sum_{k=0}^{d-1} \alpha_k \mathbbm{P}_k,
		\label{eq:satwap}
	\end{equation}
	where $\mathbbm{P}_k := \sum_{i=1}^m p(A_i - B_i = k) + p(B_i - A_{i+1} = k)$, with the convention that $A_{m+1} = A_1 + 1$ and the arithmetic is taken \textit{modulo} $d$. 
    For the CGLMP we have that $\alpha_k = 1 -2k/(d-1)$, whereas the $\alpha_k$ in the SATWAP can be tuned such that their maximal violation is achieved by a family of partially entangled states including the maximally entangled pair.
	
\begin{figure}[h!]
		\centering
		\includegraphics{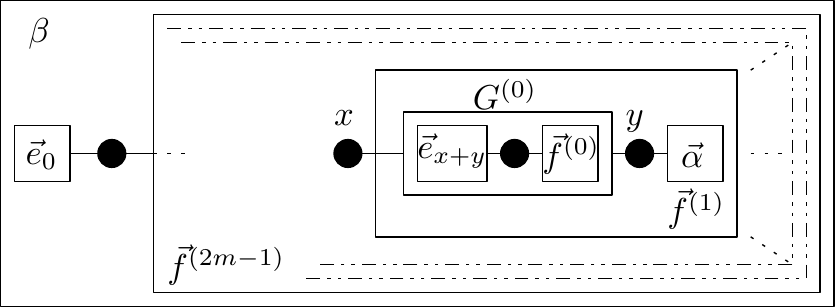}
		\caption{A graphical representation of \cref{eq:recursion}. $f^{(0)}$ is tropically contracted with the tropical canonical basis vector $\vec{e}_{x+y}$, which has a $0$ at the $(x+y)$th coordinate, $\infty$ elsewhere. This yields $G^{(0)}(x,y)$.}
		\label{fig:recursion}
	\end{figure}

	Finding out the minimum of \cref{eq:satwap} it is equivalent, by Fine's theorem~\cite{FinePRL1982}, to assigning a deterministic outcome to each of the $A_i$ and $B_i$. 
    For our purposes, it is desirable to actually make a change of variables and denote such deterministic assignment by $q_{2i}:= A_i - B_i$ and $q_{2i+1}:=B_i - A_{i+1}$, with $q_i \in [d]$. Then, we can define a cost function
	\begin{equation}
		I(\mathbf{q}) = \sum_{i\in [m]} \alpha_{q_{2i}} + \alpha_{q_{2i+1}},
	\end{equation}
	where the indices obey the redundancy condition $\sum_{i \in [2m]} q_i = 1$. 
    Hence, we can split the minimization over $\mathbf{q}$ as
	\begin{equation}
		\beta = \min_{q_0} \alpha_{q_0} + \min_{q_1} \alpha_{q_1} + \ldots + \min_{q_{2m-2}} \alpha_{q_{2m-2}} + \alpha_{1- q_0 - \ldots - q_{2m-2}}.
		\label{eq:satwapcb}
	\end{equation}
	\cref{eq:satwapcb} clearly hints a recursion taking place. 
    To make it explicit, let us define
	\begin{alignat}{5}
		f^{(0)}(x) &=& \alpha_{1-x}&,&\\
		f^{(k)}(x) &=& \min_{y} \alpha_y + f^{(k-1)}(x+y)&, &\quad 0 < k <2m.
		\label{eq:recursion}
	\end{alignat}
	Then, $\beta = f^{(2m-1)}(0)$ (see \cref{fig:recursion}). 
    In tropical terms, we define a vector $\vec{\alpha}^T = (\alpha_0, \ldots, \alpha_{d-1})$ and a Hankel circulant $2-$tensor
	\begin{equation}
		G^{(k)}= \left(
		\begin{array}{cccc}
			|&|&&|\\
			\vec{f}^{(k)}&S\vec{f}^{(k)}&\cdots&S^{d-1}\vec{f}^{(k)}\\
			|&|&&|
		\end{array}
		\right)
	\end{equation}
	where $S$ is the shift operator $S\ket{i} = \ket{i+1}$ and
	\begin{equation}
		\vec{f}^{(k)} = G^{(k-1)} \odot \vec{\alpha}^T,
	\end{equation}
	with the initial condition $f^{(0)}(x) = \alpha_{1-x}$. Then, $\beta = f^{(2m-1)}(0) = (0, \infty, \ldots, \infty)\odot \vec{f}^{(2m-1)}$.
	In \cref{fig:recursion}, we diagrammatically present the recursive tropical contraction to obtain $\beta$.
	
	\paragraph{Discussion} ---
    Tropical algebra has been rediscovered over the years in an ever-growing number of disciplines, from mathematics to physics to computer science~\cite{MaclaganBook2015}. 
    In this work, we started exploring its connection with Bell nonlocality. 
    We showed that it provides a natural framework to study some inequalities in the multipartite regime~\cite{TIpaper, TuraPRX2017, WangPRL2017, WangPRSA2018}, when the correlators are few-body and the underlying connectivity corresponds to a graph that admits an efficient contraction order, as well as certain families of inequalities for multiple outcomes~\cite{CollinsPRL2002, SATWAP}. 
    We also showed that certain renormalization flows naturally lead to the classical bound for infinite systems, in particular, those featured in~\cite{WangPRL2017}.
          
    Our study opens many questions, which we explored in detail in a parallel article~\cite{Hu2024Characterizing}. 
    In particular, we showed the relationship between the tropical eigenvectors of 
    $F$ and the facet structure of the local polytope, and their characterization as closed paths on the De Buijn graph~\cite{good_normal_1946,de_bruijn_combinatorial_1946}, which correspond to the notion of irreducible domino loops introduced in~\cite{WangPRL2017}.
    For further research, the study of multipartite nonlocality in systems with e.g. hyperbolic or fractal geometries, which can arise in the context of symmetry-protected topological order ~\cite{PhDStephen}, should be considered. 
    On the numerical side, index slicing and heuristic tensor network contraction orders can improve overall efficiency~\cite{HuangNatComputSci2021}. 
    Interestingly, the formalism here introduced captures the algorithms based on dynamic programming used for approximating the global optimum of highly nonconvex functions, such as the minimum energy of a matrix product state of a given bond dimension~\cite{SchuchPRA2010, AharonovPRA2010}, thereby providing upper bounds on the performance of any low-depth quantum circuit in the context of energy minimization.
    Our framework provides an effective toolset for the study of Bell nonlocality in multipartite quantum systems under physical geometries, enhancing its detection capabilities~\cite{Li2024Improved}, in platforms such as superconducting circuits~\cite{Wang2024Probing}.
		
	\section{Acknowledgments}
	We thank Sirui Lu and Patrick Emonts for enlightening discussions. 
    This work has received support from the European Union's Horizon Europe program through the ERC StG FINE-TEA-SQUAD (Grant No. 101040729).  
    This publication is part of the `Quantum Inspire – the Dutch Quantum Computer in the Cloud' project (with project number [NWA.1292.19.194]) of the NWA research program `Research on Routes by Consortia (ORC)', which is funded by the Netherlands Organization for Scientific Research (NWO).
	\bibliography{mylib}
	
	\end{document}